\documentclass[12pt]{article}

\newcommand{\nc}{\newcommand}

\usepackage{hyperref}

\nc{\hepref}[1]{\href{http://arXiv.org/abs/#1.html}{#1}}
\nc{\kekref}[1]{\href{http://www-lib.kek.jp/cgi-bin/img_index?#1}{KEK-19#1}}

\usepackage{mcite}
\usepackage{graphicx}
\usepackage{amsfonts}
\usepackage{amsmath}

\graphicspath{{plots/}}

\nc{\lspace}{\;\;\;\;\;\;\;\;\;\;}  \nc{\llspace}{\lspace \lspace}

\nc{\beq}{\begin{equation}}   \nc{\eeq}{\end{equation}}
\nc{\bea}{\begin{eqnarray}}   \nc{\eea}{\end{eqnarray}}
\nc{\bi}{\begin{itemize}}    \nc{\ei}{\end{itemize}}
\nc{\be}{\begin{enumerate}}  \nc{\ee}{\end{enumerate}}
\nc{\bc}{\begin{center}}     \nc{\ec}{\end{center}}
\nc{\ba}[1]{\begin{array}{#1}}     \nc{\ea}{\end{array}}
\nc{\macierz}[2]{\left(\begin{array}{#1} #2 \end{array}\right)}

\nc{\vts}{\mkern1mu}
\nc{\ri}{\mathrm{i}\vts}
\nc{\sw}{\sin\theta_W}
\nc{\cw}{\cos\theta_W}

\nc{\tb}{\tan\beta}

\nc{\tr}{\mathrm{Tr}}

\nc{\dslash}[1]{\slash\!\!\!#1}

\nc{\CPV}{CPV}

\nc{\Refrmfl}{\Re(f_2^R-\bar{f}_2^L)}
\nc{\ReDZ}{\Re D_Z}
\nc{\ReDgamma}{\Re D_{\gamma}}

\nc{\nnr}{\nonumber}

\nc{\jpXchecked}{{\Large Formulas X-checked\newline}}

\nc{\jpfigure}[4]{%
  \begin{figure}%
  \begin{center}%
  \includegraphics[#1]{#2}%
  \end{center}%
  \caption{#3}%
  \label{#4}%
  \end{figure}
}%

\nc{\loopC}[2]{{\mathbf{C}}_{#1}^{#2}}

\nc{\fdiag}{0}
\nc{\bg}{B. Grzadkowski}
\nc{\BG}{Bohdan Grzadkowski}
\nc{\lsp}{\;\;\;\;\;\;\;\;}
\nc{\non}{\nonumber}
\nc{\lumun}{\;{\hbox {fb}^{-1}}{\hbox {y}^{-1}}}
\nc{\hc}{\hbox {h.c.}}
\nc{\re}{\hbox {Re}}
\nc{\im}{\hbox {Im}}
\nc{\etal}{\hbox{et al.}}
\nc{\pbarn}{\;\hbox {pb}}
\nc{\ra} {\rightarrow}
\nc{\ctw}{\cos\theta_W} 
\nc{\stw}{\sin\theta_W}
\nc{\ctwsq}{\cos^2\theta_W}        
\nc{\stwsq}{\sin^2\theta_W}
\nc{\ttbar}{t\bar{t}}
\nc{\bbbar}{b\bar{b}}
\nc{\tanb} {\tan \beta}
\nc{\twbdec} {t\rightarrow W^+ b}
\nc{\tbwbdec} {\bar{t} \rightarrow W^- \bar{b}}
\nc{\hprod} {e^+e^- \ra Z^\ast \ra h Z}
\nc{\epem} {e^+e^-}
\nc{\wpwm} {W^+W^-}
\nc{\tbar} {\bar{t}}
\nc{\bbar} {\bar{b}}
\nc{\wpp} {W^+}
\nc{\mt}{m_t}
\nc{\mts}{m_t^2}
\nc{\mw} {m_W}
\nc{\mws} {m_W^2}
\nc{\mz} {m_Z}
\nc{\mzs} {m_Z^2}
\nc{\mh} {m_h}
\nc{\mhs} {m_h^2}
\nc{\ma} {m_A}
\nc{\mas} {m_A^2}
\nc{\hdec}{h \ra t\bar{t}}
\nc{\ttbardec}{\ttbar \ra W^+W^-\bbbar}
\nc{\po}{\Phi_1}
\nc{\pht}{\Phi_2}
\nc{\phtd}{\Phi_2^\dagger}
\nc{\phtt}{{\tilde{\Phi}}_2}
\nc{\popo}{\po^\dagger\po}
\nc{\phtpt}{\pht^\dagger\pht}
\nc{\popt}{\po^\dagger\pht}
\nc{\phtpo}{\pht^\dagger\po}
\nc{\sq}{\sqrt{2}}
\nc{\nsd} {N_{SD}}
\nc{\ntt} {N_{tt}}
\nc{\vs}{\vspace{2mm}}
\nc{\sty}{\hat{S}^t_1} \nc{\pty}{\hat{P}^t_1}
\nc{\sts}{(\sty)^2}      \nc{\pts}{(\pty)^2}
\nc{\yts}{\sts+\pts}
\nc{\sby}{\hat{S}^b_1} \nc{\pby}{\hat{P}^b_1}
\nc{\sbs}{(\sby)^2}      \nc{\pbs}{(\pby)^2}
\nc{\ybs}{\sbs+\pbs}
\nc{\eettz}{\epem \rta \ttbar Z}
\nc{\barx}{\bar{x}}
\nc{\bb}{\stackrel{{\scriptscriptstyle (-)}}{b}}

\nc{\et}{\tilde{e}}
\nc{\ft}{\tilde{f}}
\nc{\gt}{\tilde{g}}
\nc{\hti}{\tilde{h}}

\def\sb{s_\beta}
\def\cb{c_\beta}

\def\lsim{\mathrel{\raise.3ex\hbox{$<$\kern-.75em\lower1ex\hbox{$\sim$}}}}
\def\gsim{\mathrel{\raise.3ex\hbox{$>$\kern-.75em\lower1ex\hbox{$\sim$}}}}

\def\mev{\,{\rm MeV}}
\def\gev{\,{\rm GeV}}

\def\rta{\rightarrow}

\def\dps{\displaystyle}
\def\thf{\theta_{\!\scriptscriptstyle{f}}}
\def\sst#1{\scriptscriptstyle{#1}}
\def\ssf{\scriptscriptstyle{f}}

\begin{document}
%\begin{flushright}
%\today
%\end{flushright}
%
\font\fortssbx=cmssbx10 scaled \magstep2
\medskip
\begin{flushright}
$\vcenter{
\hbox{\bf IFT-30/2000}
\hbox{December, 2000}
}$
\end{flushright}
\vspace*{1.5cm}
\begin{center}
{\large{\bf Testing Scalar-Sector CP Violation in Top-Quark Production and Decay at Linear {$\mathbf \epem $} Colliders}}\\ 
\rm
\vspace*{1cm}

{\bf \BG}\footnote{E-mail:{\tt bohdang@fuw.edu.pl}}  and
{\bf Jacek Pliszka}\footnote{E-mail:{\tt pliszka@fuw.edu.pl}}\\ 

\vspace*{1cm}
{\em Institute of Theoretical Physics, Warsaw University, Warsaw, Poland}\\

\vspace*{2cm}

{\bf Abstract}
\end{center}
\vspace{5mm} 
We consider a general two-Higgs-doublet model with CP violation in the scalar sector.
Three neutral Higgs fields of the model all mix and the resulting physical Higgs bosons
have no definite CP properties. That leads, at the one-loop level of the perturbation expansion,
to CP-violating form factors for  $\gamma\ttbar$, $Z\ttbar$ and $Wtb$ interaction vertices.
We discuss asymmetries sensitive to CP violation induced by the form factors for the process
$\epem \to \ttbar \to l^\pm \cdots$ and $\epem \to \ttbar \to \bb \cdots$ at future linear 
$\epem$ colliders.

\vfill
\setcounter{page}{0}
\thispagestyle{empty}
\newpage

%--------------------------------------------------------------------
%\renewcommand{\thefootnote}{\sharp\arabic{footnote}}
%\setcounter{footnote}{0}
%--------------------------------------------------------------------

%%%%%%%%%%%%%%%%%%%%%%%%%%%%%%%%%%%%%%%%%%%%%%%%%%%

\section{Introduction}

Even though the top quark has been already discovered  
several years ago~\cite{Abe:1994xt,*Abachi:1995iq}, 
its~interactions are still weakly constrained. It remains an open 
question if top-quark couplings obey the Standard Model (SM) 
scheme of the electroweak forces or there exists a contribution from 
physics beyond the SM. In particular, CP violation in the top-quark
interactions has not been verified. 
The classical method for
incorporating CP violation into the SM is to make 
the Yukawa couplings of the Higgs boson to
quarks explicitly complex,  as  built into the Kobayashi-Maskawa
mixing matrix~\cite{Kobayashi:1973fv} proposed more than two decades ago. 
However, CP violation could equally well be partially or wholly due
to other mechanisms. The possibility that 
CP violation derives largely from the Higgs sector 
itself is particularly appealing in the context of the observed baryon asymmetry,
since its explanation requires more CP violation~\cite{Gavela:1994dt,*Huet:1995jb} 
then is provided by the SM. Even 
the simple two-Higgs-doublet model  (2HDM) extension of the 
one-doublet SM Higgs sector
provides a much richer framework for describing CP violation; in the 2HDM,
spontaneous and/or explicit CP violation is possible
in the scalar sector~\cite{Lee:1973iz,*Weinberg:1990me,*Branco:1985aq}.
The model, besides CP violation, offers many other appealing phenomena, for 
a review see Ref.~\cite{HHG}. 

For our analysis, the most relevant part of the interaction Lagrangian 
takes the following form~\footnote{One could also consider more general, CP-violating $ZZh$ 
coupling, see Ref.~\cite{Han:2000mi}, however here the contribution from such a vertex
would be negligible.}:
\beq
{\cal{L}}= -\frac{\mt}{v}h\tbar(a+i\gamma_5 b)t + 
C \frac{h}{v} (\mz^2 Z_\mu Z^\mu+2 \mw^2 W_\mu W^\mu),
\label{lag}
\end{equation}
where $h$ is the lowest mass scalar, $g$ is the SU(2) coupling constant, 
$v$ is the Higgs boson vacuum expectation value 
(with the normalization adopted here such that $v=2m_W/g=246\,\gev$), 
$a$, $b$ and $C$ are real parameters which account for deviations from the
SM, $a=1$, $b=0$ and $C=1$ reproduce the SM Lagrangian. 
Since under CP, $\tbar(a+i\gamma_5 b)t \stackrel{{\rm CP}}{\ra} \tbar(a-i\gamma_5 b)t$ and 
$Z_\mu Z^\mu  \stackrel{{\rm CP}}{\ra} Z^\mu Z_\mu$, one can observe that 
terms in the cross section proportional to $ab$ or $bC$ would indicate 
CP violation. 
The~2HDM is the minimal extension of the SM that provides non-zero $ab$ and/or~$bC$.

In this paper we will focus on CP-violating contributions to the process 
$\epem \to \ttbar \to l^\pm  \cdots$ and $\epem \to \ttbar \to \bb \cdots$ 
induced within 2HDM. However the fundamental goal is seeking for the
ultimate theory of electroweak interactions. There are several reasons to utilize CP violation
in the top physics while looking for physics beyond the SM:
\bi
\item The top quark decays immediately after being produced as its huge
mass $m_t=174.0\pm3.2\pm4.0\gev$~\cite{Groom:2000in}
leads to a decay width ${\mit\Gamma}_t$ much
larger than~${\mit\Lambda}_{\rm QCD}$. 
Therefore the decay process is
not contaminated by any fragmentation 
effects~\cite{Bigi:1981az,*Kuhn:1982ua,*Bigi:1986jk} and decay
products may provide useful information on top-quark properties.
\item Since the top quark is heavy,
its Yukawa coupling is large and therefore its interactions could be 
sensitive to a Higgs sector of the electroweak theory. 
\item At the same time, the TESLA collider design is supposed to offer an integrated
luminosity of the order of $L=500\lumun$ at $\sqrt{s}=500\gev$. 
Therefore expected number of $\ttbar$ events per year could reach $5\times10^4$ 
even for $\ttbar$ tagging efficiency $\epsilon_{t\bar{t}}=15\%$. 
That should allow to study subtle properties of the top quark, which could
e.g. lead to CP-sensitive asymmetries of the order of~$5\times 10^{-3}$.
\item Since the top quark is that heavy  and the third family of quarks effectively 
decouples from the first two,
any CP-violating observables within the SM are expected to be tiny, e.g.: {\it i)} non-zero electric dipole 
moment of fermions is generated at the three-loop approximation of the perturbation 
expansion~\cite{Czarnecki:1997bu}, or {\it ii)} the decay rate asymmetry (being a one-loop effect) 
is strongly GIM suppressed reaching at most a value $10^{-9}$~\cite{Grzadkowski:1993gh}.
So, one can expect that for CP-violating asymmetries any SM background could be safely
neglected.
\ei
Therefore it seems to be justified to look 
for CP-violating Higgs effects in the process of $\ttbar$ production and its subsequent decay at future
linear $\epem$ colliders. Even though 2HDM contributions to various CP-sensitive asymmetries
has been already presented in the existing 
literature, see Refs.~\cite{Chang:1993fu,*Bernreuther:1992dz,Grzadkowski:1992yz}, 
here we are providing a consistent
treatment of CP violation both in the production, $\epem \to \ttbar$,  {\it and} in the top-quark decay,
$t\to b W$.  For an extensive review of CP violation in top-quark interactions see 
Ref.~\cite{Atwood:2000tu}.

The paper is organized as follows. In Section~\ref{model}, we briefly outline the mechanism of CP 
violation in the 2HDM, introduce the mixing matrix for neutral scalars and derive necessary couplings.
In section~\ref{ffactors},  we present results for CP-violating form factors both 
for the $\ttbar$ production
process and for $t$ and $\bar{t}$ decays.  
In Section~\ref{expcon}, we recall current experimental constraints
relevant for the CP-violating observables considered in this paper.
In Section~\ref{asymm}, we collect results for various energy and angular CP-violating asymmetries.
Concluding remarks are given in Section~\ref{summary}.

%%%%%%%%%%%%%%%%%%%%%%%%%%%%%%%%%%%%%%%%%%%%%%%%%%%

\section{The two-Higgs-doublet model with CP violation}
\label{model}

The 2HDM of electroweak interactions 
 contains two SU(2) Higgs doublets denoted by 
$\Phi_1=(\phi_1^+,\phi_1^0)$ and $\Phi_2=(\phi_2^+,\phi_2^0)$.
It is well known~\cite{Lee:1973iz,*Weinberg:1990me,*Branco:1985aq} that the model 
allows both for spontaneous and explicit CP violation\footnote{Here we are considering
a model with discrete $Z_2$ symmetry that prohibits flavor changing neutral 
currents. In order to
allow for CP violation the symmetry has to be broken softly by the term 
$\mu_{12}^2\Phi_1^\dagger\Phi_2$ in the potential.}.

After
SU(2)$\times$U(1) gauge symmetry breaking, one combination of neutral
Higgs fields, $\sqrt2(\cb\Im\phi_1^0+ \sb\Im\phi_2^0)$,
becomes a would-be Goldstone boson which is absorbed while giving 
mass to the $Z$ gauge boson.
(Here, we use the notation $\sb\equiv\sin\beta$, $\cb\equiv\cos\beta$,
where $\tanb=\langle \phi_2^0 \rangle/\langle \phi_1^0 \rangle$.)
The same mixing angle, $\beta$, also diagonalizes 
the mass matrix in the charged Higgs sector.  
If either explicit or spontaneous CP violation is present,
the remaining three neutral degrees of freedom, 
\begin{equation}
(\varphi_1,\varphi_2,\varphi_3)\equiv
\sqrt 2(\Re\phi_1^0, \, \Re\phi_2^0, \,
 s_\beta\Im\phi_1^0-c_\beta\Im\phi_2^0) 
\end{equation} 
are not mass eigenstates. The physical neutral Higgs bosons $h_i$
($i=1,2,3$) are obtained by an orthogonal transformation, $h=R
\varphi$, where the rotation matrix is given in terms of three Euler
angles ($\alpha_1, \alpha_2,\alpha_3$) by
\begin{eqnarray} 
R=\left(\ba{ccc}
  c_1     &  -s_1c_2          &     s_1s_2  \\
  s_1c_3  & c_1c_2c_3-s_2s_3  &  -c_1s_2c_3-c_2s_3\\
  s_1s_3 & c_1c_2s_3+s_2c_3 & -c_1s_2s_3+c_2c_3 \ea\right),
\label{mixing}
\end{eqnarray}
where $s_i\equiv\sin\alpha_i$ and $c_i\equiv\cos\alpha_i$.

As a result of the mixing between real and imaginary parts of neutral
Higgs fields, the Yukawa interactions of the $h_i$ mass-eigenstates are not
invariant under CP. They are given by:
\begin{equation} 
{\cal L}=-\frac{m_f}{v} h_i\bar{f}(a^f_i+ib^f_i\gamma_5)f \label{coupl} 
\end{equation}
where the scalar ($a^f_i$) and pseudoscalar ($b^f_i$) couplings are
functions of the mixing angles. For up-type quarks we have 
\begin{equation}
a^u_i=\frac{1}{s_\beta}R_{i2},\;\;\;\;\;
b^u_i=\frac{c_\beta }{s_\beta}R_{i3}, 
\label{absu}
\end{equation}
and for down-type quarks:
\begin{equation}
a^d_i=\frac{1}{c_\beta}R_{i1},\;\;\;\;\;
b^d_i=\frac{s_\beta}{c_\beta} R_{i3}\,,
\label{absd}
\end{equation}
and similarly for charged leptons. 
For large $\tan\beta$, the couplings to down-type fermions are 
typically enhanced over the couplings to up-type fermions.

In the following analysis we will also need  
the couplings of neutral Higgs and vector bosons, they are given by
\begin{equation}
g_{VVh_i} \equiv 2 \frac{ m_V^2}{v} C_i= 
2 \frac{ m_V^2}{v} (s_{\beta} R_{i2}+c_{\beta}R_{i1}),
\label{zzcoups}
\end{equation}
for $V=Z,W$.
Hereafter we shall denote the lightest Higgs boson by $h$ and its $R$-matrix 
index by~$i$.

%%%%%%%%%%%%%%%%%%%%%%%%%%%%%%%%%%%%%%%%%%%%%%%%%%%%%%%%%%%%

\section{Form Factors}
\label{ffactors}
\subsection{{\boldmath $\ttbar$} Production}

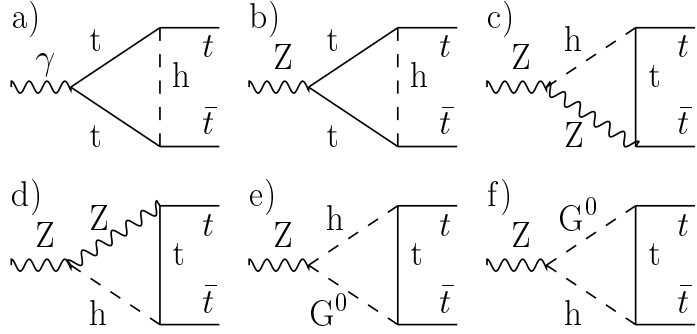
\jpfigure{width=0.65\textwidth}{phd-diag-eett6.eps}{Diagrams contributing to CP-violating form factors
$D_{\gamma,Z}$.}{fig:diag:eett}

The effective $\ttbar \gamma$ and $\ttbar Z$  vertices will be parameterized by the following form 
factors\footnote{Two other possible
    form factors do not contribute in the limit of zero electron 
    mass.}:

\begin{equation}
{\mit\Gamma}^\mu_v=\frac{g}{2}\bar{u}(p_t)
\biggl[\,\gamma^\mu(A_v-B_v\gamma_5)
+\frac{(p_t-p_{\bar{t}})^\mu}{2\mt}(C_v-D_v\gamma_5)\,\biggr]v(p_{\bar{t}}),
\label{vtt}
\end{equation}
where $g$ denotes the SU(2) gauge coupling constant, $v=\gamma,Z$,
and 
$$
A_\gamma=-\frac43\sw,\ \ B_\gamma=0,\ \ 
A_Z=-\frac{v_t}{2\cw},\ \ 
B_Z=-\frac{a_t}{2\cw},\ \ 
C/D_{\gamma,Z}=0
$$
denote the SM contributions to the vertices for
$$
v_t=\Bigl(1-\frac83\sin^2\theta_W\Bigr) \lsp
a_t=1.
$$
The 
form factors $A_v$, $B_v$, $C_v$ describe
$CP$-conserving while $D_v$ parameterizes $CP$-violating
contributions. 

Further in this paper the following parameters will be adopted:
$m_t=175\gev$, $m_Z=91.187\gev$, 
$\Gamma_Z=2.49\gev$, $\sin^2 \theta_W=0.23$ and $m_b=4.2\gev$.

Since in this paper we are focusing on CP-violating asymmetries, 
the only relevant form factors are 
$D_\gamma$ and $D_Z$. Direct calculation of diagrams shown in Fig.\ref{fig:diag:eett} leads to 
the following result in terms of 3-point Passarino-Veltman~\cite{PassarinoVeltman}  
functions defined in the appendix \ref{app-loops}: 
\bea
\label{eqn:eett:dgammadz}
D_\gamma&=&
\frac{\ri}{2\pi^2} A_\gamma
\frac{m_t^2}{v^2} 
b_i^t a_i^t  m_t^2 
\loopC{12}{}(p_t, p_{\bar{t}}, m_t^2, m_h^2, m_t^2),\nnr\\
D_Z&=& 
\frac{\ri}{2\pi^2} A_Z  \frac{m_t^2}{v^2}
b_i^t \left[
a_i^t m_t^2 
\loopC{12}{}(p_t,  p_{\bar{t}}, m_t^2, m_h^2, m_t^2)\right.\nnr\\
&& 
\left. -  
C_i m_Z^2 
\loopC{12}{}(p_t,p_{\bar{t}}, m_h^2, m_t^2, m_Z^2) \right].
\eea
Since the asymmetries we are going to discuss are generated by real parts of the above form factors,
let us decompose (using eqs.(\ref{absu}, \ref{zzcoups})) $\Re(D_\gamma)$ and $\Re(D_Z)$
in the following way\footnote{The formulae for $\Re D_{\gamma}$ and $\Re D_Z$ 
confirm results published in Ref.~\cite{Chang:1993fu,*Bernreuther:1992dz}.}:
\bea
\Re D_{\gamma}&=& R_{i2} R_{i3} f_{23}^{\gamma (a)}\\
\Re D_Z&=& R_{i2} R_{i3} (f_{23}^{Z(b)}+f_{23}^{Z(c-f)})+R_{i1} R_{i3} f_{13}^{Z(c-f)},\non 
\label{sec:eett:prod:plots:fgdefs}
\end{eqnarray}
where superscripts indicate graphs that generate the contribution according to the notation of 
Fig.\ref{fig:diag:eett}.
Since in the case of the photon only Yukawa couplings $a_i^t$ and $b_i^t$ contribute, any signal of 
CP violation must be proportional to $a_i^t b_i^t \sim R_{i2} R_{i3}$. However, for the $Z$ boson 
vertex there is also other ``source'' of CP violation, namely $b_i^t  C_i \ni R_{i1} R_{i3}$.

It is useful to discuss $\tanb$ dependence of the  functions $f$ first.
From eq.(\ref{eqn:eett:dgammadz}) and  eqs.(\ref{absu}, \ref{zzcoups}) one can find out 
that all contributions to the form factors $D_{\gamma}$, $D_Z$ are enhanced for  small $\tanb$:
$f_{23}^{\gamma (a)},\; f_{23}^{Z(b)} \sim \tan^{-2}\beta$,   $f_{23}^{Z(c-f)} \sim \cos\beta$ and
$f_{13}^{Z(c-f)} \sim \tan^{-1}\beta$. Therefore, for $\tanb < 1$, 
the contributions from diagrams c)-f) are expected 
to be suppressed relatively to those generated by diagrams a) or b) in Fig.\ref{fig:diag:eett}.

Hereafter we assume that there exists only one light Higgs boson $h$ and 
possible effects of 
the heavier scalar degrees of freedom decouple.
In Fig.\ref{fig:eett:prodre:hdep} we illustrate dependence of 
the functions $f$ on the 
lightest Higgs boson mass, $m_h$.
In order to amplify possible contributions we have chosen  $\tanb=0.5$. 
As it is seen from the figure, the dominant CP-violating effects
will be generated by $f_{23}^{\gamma (a)}$ and $f_{23}^{Z (b)}$, which are generated by
the Yukawa-coupling contribution, $a_i^t b_i^t$, from diagrams a) and b) in Fig.\ref{fig:diag:eett}.
As one could have anticipated, we observe an enhancement of the contributions 
generated by the diagrams a) and b) for
low Higgs boson mass. The growth of $f_{23}^{\gamma (a)}$ and $f_{23}^{Z (b)}$ is more pronounced
for lower $\sqrt{s}$ (closer to the $\ttbar$ production threshold), it is a consequence of the Coulomb-like 
singularity generated by graph a) and b) in the limit $m_h \to 0$ at the $\ttbar$ production threshold. 
Similar behavior have been also noted 
in the case of CP-conserving form factors~\cite{Guth:1992ab}.
One also observes a typical threshold behavior at $(\sqrt{s}-m_Z)$
for $f_{23}^{Z(c-f)}$ and  $f_{13}^{Z(c-f)}$ as a 
non-trivial absorptive part of diagrams c)-f) is necessary for nonzero contribution to 
$\Re D_Z$\footnote{The same applies for $\Re D_\gamma$.}.

If all three Higgs bosons of the model have the same mass, then the
orthogonality of the mixing matrix $R_{ij}$ guarantees~\cite{Branco:1999fs} vanishing CP violation
while summed over all the scalars.
Therefore, the leading contribution
to $\Re D_\gamma$ and $\Re D_Z$ originating from an exchange of the lightest Higgs boson $h$
could be partially cancelled by heavier scalars $h_h$. 
However here we will assume that masses of the heavier scalars are above 
the production threshold for $Zh_h$, therefore, as observed (for the lightest Higgs boson mass
below $100\gev$)
from Fig.\ref{fig:eett:prodre:hdep}, 
the cancellation by heavier scalars for $\sqrt{s}=360,\;,500$ and $1000\gev$ could reach
at most $20,\;30$ and $40\%$\footnote{Thus the cancellation expected in our case is not
that strong as obtained in Ref.~\cite{Hasuike:1998pn,*Hasuike:1996tr} 
for a specific choice of model parameters.}.
 
Leading asymptotic formulae for small and large Higgs mass are presented in the 
appendix~\ref{app-asymptotic}. It is worth to notice that for non-zero $\beta_t=\sqrt{1-4\mt^2/s}$,
in the limit $m_h^2/(\beta^2_t s) \to 0$ both $\Re D_{\gamma}$ and $\Re D_Z$ are finite,
whereas a typical decoupling limit, $\Re D_{\gamma}, \Re D_Z \sim m_t^4/(m_h^2m_W^2)$, is observed 
for large $m_h$.

Fig.\ref{fig:eett:prodre:sdep} shows the functions $f$ for $\tanb=0.5$ and two different
Higgs boson masses $m_h=10\gev$ and $m_h=100\gev$, 
as a function of $\sqrt{s}$. It is seen that it is not desirable to choose too high beam energy, 
as the size of the functions drops. Again, $f_{23}^{\gamma (a)},\; f_{23}^{Z(b)}$ dominate 
over $f_{13}^{Z(c-f)},\; f_{23}^{Z(c-f)}$  by 1-2 orders of magnitude.

\jpfigure{width=.65\textwidth}{plot-eett-prodre-ffmhdep.eps}{
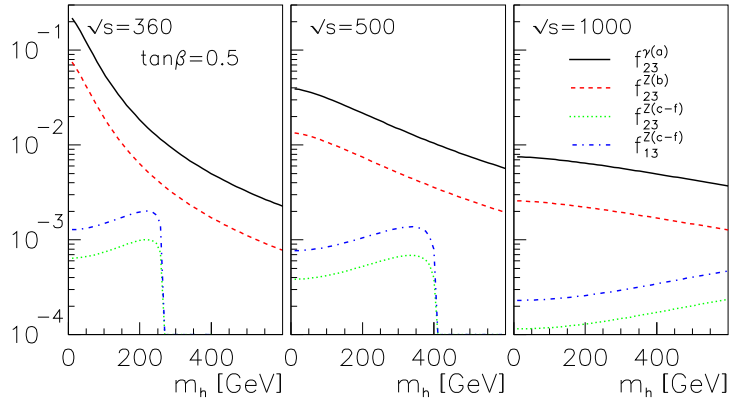
The functions $f$ defined in eq. (\ref{sec:eett:prod:plots:fgdefs}) as a function of $m_h$ for
$\tanb=0.5$, $\sqrt{s}=360$ (left), $\sqrt{s}=500$ (middle)  and $\sqrt{s}=1000\gev$ 
(right).}{fig:eett:prodre:hdep}

\jpfigure{width=.65\textwidth}{plot-eett-prodre-ffsdep.eps}{
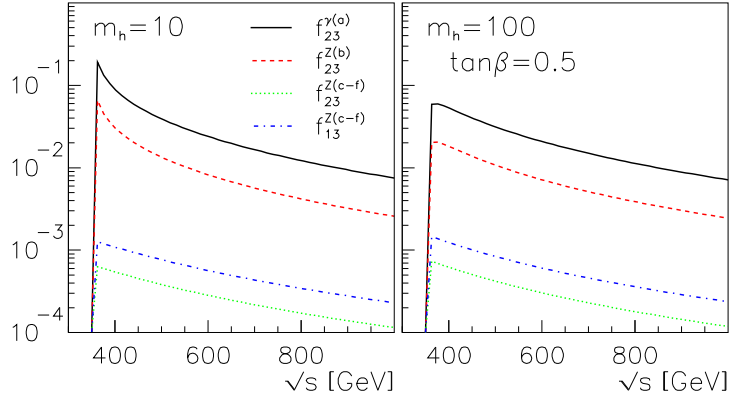
The functions $f$ defined in eq. (\ref{sec:eett:prod:plots:fgdefs}) as a function of  $\sqrt{s}$ for
$\tanb=0.5$, $m_h=10\gev$ and $m_h=100\gev$}{fig:eett:prodre:sdep}

\subsection{Top Decay}

We will adopt the following parameterization of
the $Wtb$ vertex suitable for the $t$ and $\tbar$ decays:
\bea
{\Gamma}^{\mu}&=&-\frac{g}{\sqrt{2}}V_{tb}\:
\biggl[\,\gamma^{\mu}(f_1^L P_L +f_1^R P_R)
-\frac{{i\sigma^{\mu\nu}k_{\nu}}}{M_W}
(f_2^L P_L +f_2^R P_R)\,\biggr],\ \ \ \ \non \\
\bar{\Gamma}^{\mu}&=&-\frac{g}{\sqrt{2}}V_{tb}^*\:
\biggl[\,\gamma^{\mu}(\bar{f}_1^L P_L +\bar{f}_1^R P_R)
-\frac{{i\sigma^{\mu\nu}k_{\nu}}}{M_W}
(\bar{f}_2^L P_L +\bar{f}_2^R P_R)\,\biggr],
\label{tbw}
\eea
where $P_{L/R}=(1\mp\gamma_5)/2$, $V_{tb}$ is the $(tb)$ element of
the Kobayashi-Maskawa matrix and $k$ is the momentum of $W$. 
In the SM $f_1^L=\bar{f}_1^L=1$ and all the other form factors vanish.
It turns out that in the limit of massless bottom quarks the only form factors that interfere with
the SM are $f_2^R$ and $\bar{f}_2^L$ for the top and anti-top decays, respectively.
Currently, there is no relevant 
experimental bound on those form factors\footnote{There exists direct experimental constraints from the Fermilab Tevatron on the form factors
that are obtained through the determination of the $W$-boson helicity. Pure $V-A$
theory for massless bottom quarks predicts an absence of positive helicity
$W^+$ bosons, therefore the upper limit on the helicity ${\cal F}_+$ implies
an upper limit on the $V+A$ coupling $f_1^R$, however, the resulting  
limit is rather weak~\cite{Affolder:1999mp}. There exist an indirect, but much stronger 
bound~\cite{Cho:1994zb,*Fujikawa:1994zu} on
the admixture of right-handed currents, $\bar{f}_1^R$, coming from data for 
$b \to s \gamma$, namely $-0.05\lsim\bar{f}_1^R\lsim0.01$.}.

One can show that the CP-violating and CP-conserving parts of the
form factors for $t$ and $\tbar$ are not independent:
\beq
f_1^{L,R}=\pm\bar{f}_1^{L,R} \;\;{\rm and} \;\;\;f_2^{L,R}=\pm\bar{f}_2^{R,L},
\label{cp_relation}
\end{equation}
where upper (lower) signs are those for $CP$-conserving
(-violating) contributions~\cite{Bernreuther:1992be,Grzadkowski:1992yz}. Therefore any
$CP$-violating observable defined for the top-quark decay must be
proportional to $f_1^{L,R}-\bar{f}_1^{L,R}$ or $f_2^{L,R}-
\bar{f}_2^{R,L}$.

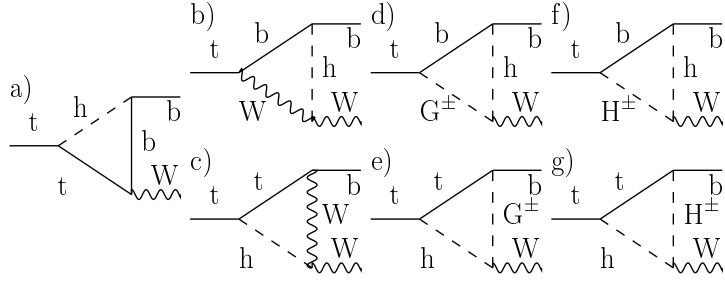
\jpfigure{width=0.65\textwidth}{phd-diag-tbw.eps}{Diagrams contributing to $f_2^R-
\bar{f}_2^L$.}{fig:diag:tbw}

Diagrams contributing to CP violation in the decay process are shown in Fig.\ref{fig:diag:tbw}.
Direct calculation leads to the following result for the CP-violating part
of $f_2^R$~\footnote{The general result for $f_2^R|_{CPV}$ agrees with  
formulae for $\Im f_2^R|_{CPV}$  from Ref.~\cite{Grzadkowski:1992yz}.}:
\begin{eqnarray}
\label{eq:eettp:f2R}
&f_2^R|_{CPV}=\frac{\ri g}{32 \pi^2}\frac{m_b}{v} m_t m_b \Big\{
  \left[ 2 a_i^b b_i^t \loopC{12}{a} + 
      \left( a_i^b b_i^t - a_i^t b_i^b \right) \loopC{23}{a} \right] 
  - b_i^b C_i \loopC{22}{bd} &\nnr\\
&  + b_i^t C_i \left[ \loopC{22}{ce}  + 
      \loopC{23}{ce}\left( \frac{m_t^2}{m_b^2}-1\right)
          + 2\left( \loopC{12}{ce} - \loopC{11}{ce} \right) 
       \frac{m_W^2}{m_b^2} \right]&\\ 
&  + \left[ 
     2 a_i^b C^2_i \loopC{12}{f} ct_\beta   
     -\loopC{22}{f} \Im(y_i^b C^-_i) t_\beta 
     + \loopC{23}{f} \left( \Im(y_i^b C^+_i) ct_\beta + 
         \Im(y_i^b C^-_i) t_\beta \right) 
      \right]&\nnr\\
&\left. 
  - \left[ 
     2 a_i^t C^2_i \loopC{12}{g} ct_\beta \frac{m_t^2}{m_b^2} 
             -\loopC{22}{g} \Im(y_i^t C^-_i) t_\beta 
             +\loopC{23}{g} \left( \Im(y_i^t C^+_i) \frac{m_t^2}{m_b^2}
              ct_\beta + 
         \Im(y_i^t C^-_i) t_\beta \right)
      \right] \right\}
&\nnr\end{eqnarray}
where \bea
\loopC{ij}{a} &=&
\loopC{ij}{}(p_t, -p_W, m_h^2, m_t^2, m_b^2)\nnr\\
\loopC{ij}{bd}&=&
\loopC{ij}{}(p_W,  p_b, m_W^2, m_h^2, m_b^2)\nnr\\
\loopC{ij}{ce}&=&
\loopC{ij}{}(p_W,  p_b, m_h^2, m_W^2, m_t^2)\nnr\\
\loopC{ij}{f} &=&
\loopC{ij}{}(p_W, -p_t, m_h^2, m_H^2, m_b^2)\nnr\\
\loopC{ij}{g} &=&
\loopC{ij}{}(p_W,  p_b, m_h^2, m_H^2, m_t^2) 
%\ \ \ \ \ \ \ \ \ \textrm{and}
\eea
and
\begin{equation}
y_i^{t/b}\equiv a_i^{t/b}+\ri b_i^{t/b},\ \ \ C_i^{\pm}\equiv C_i^{(1)} \pm \ri C_i^{(2)},
\ \ \ t_\beta=\tan\beta,\ \ ct_\beta=\cot\beta,
\end{equation}
where $C_i^{(1)}=  \sb R_{i1}-\cb R_{i2}$ and $C_i^{(2)}=R_{i3}$. 
The asymmetries that will be discussed here depends on 
real parts of $f_2^R$ and $\bar{f}_2^L$.
It is easy to note from eq.(\ref{eq:eettp:f2R}) that
imaginary parts of $\loopC{}{}$ 
functions contribute to the real part of $f_2^R|_{CPV}$.
It is seen that only diagrams b), d) and f) will contribute to $\Re(f_2^R|_{CPV})$. 
However,  $b\ra s\gamma$ strongly suggest~\cite{Greub:1999sv} 
that for 2HDM  $m_{H^{\pm}} > m_t - m_b$, therefore eventually (adopting the relation
(\ref{cp_relation})) one gets the following
result (from graphs b) and d) only) for CP-violating contribution to $\Re(f_2^R|_{CPV})$:
\begin{equation}
\Re(f_2^R-\bar{f}_2^L)=2\Re(f_2^R|_{CPV})=\frac{g }{16 \pi^2 } \frac{m_b}{v} m_b m_t
    b_i^b C_i \Im \loopC{22}{bd}
\label{f2r}
\end{equation}
As will be discussed in Section~\ref{expcon} there is a strong experimental bound on $|C_i|$ for
$m_h < 105 \gev$. Taking into account the limit on $|C_i|$ and choosing $\tan\beta=50$
(in order to illustrate a possible enhancement) we
plot $\Re(f_2^R-\bar{f}_2^L)$ in Fig.~\ref{fig:diag:tbw}  
as a function of $m_h$. It is seen that
$\Re(f_2^R-\bar{f}_2^L)$ is by $2-4$ orders of magnitude below $\Re D_\gamma$
or $\Re D_Z$ even for large b-quark Yukawa coupling,
compare Fig.~\ref{fig:eett:prodre:hdep} and~\ref{fig:eett:prodre:sdep}.
The~suppression is caused both by the experimental limit on 
$|C_i|$  (for $m_h < 105\gev$)
and by an extra suppression factor of $(m_b/m_t)^2$ 
(relative to $\Re D_{\gamma,Z}$).

There is a comment in order here; since in the 2HDM the real part of CP-violating 
form factors in the top decay
is much smaller then in the production process, it is interesting to look closer
at the
suppression mechanism and find class of possible extensions of the SM 
that provide large $\Re(f_2^R-\bar{f}_2^L)$. 
Since an absorptive part is needed, the only graphs that may contribute are those
denoted by b), d) and f) in Fig.\ref{fig:diag:tbw} (allowing for the neutral scalar to be
replaced by a neutral vector). 
One source of the suppression is the bottom-quark  mass that  
originate from the propagator while the second one comes from
the Yukawa vertex. The latter one could be easy amplified by large $\tanb$: for
$\tanb \simeq 38$ the bottom-quark Yukawa coupling is as strong as the SU(2) gauge
coupling. Therefore the suppression to overcome is $m_b$ from the bottom-quark propagator.
A possible solution~\cite{work_in_progress}
seems to be a multi-doublet-Higgs model that could evade the stringent
restriction from the $b \to s \gamma$ decay and also overcome (through a contribution from the graph 
type f) in Fig.\ref{fig:diag:tbw})  the limit 
on $W^+W^- h$ that comes from the LEP limit on $|C_i|$.
Let us notice that it is much
easier 
to develop large $\Im(f_2^R-\bar{f}_2^L)$, see e.g. 
Refs.~\cite{Grzadkowski:1992yz} and~\cite{Atwood:2000tu}. 

\jpfigure{width=0.65\textwidth}{plot-eett-decay-ffmhdep.eps}
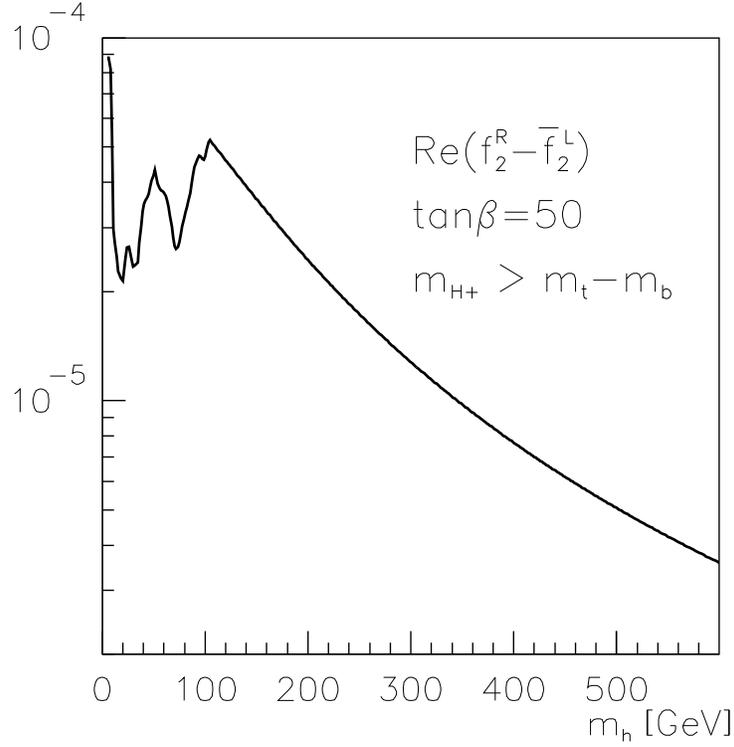
{$\Re(f_2^R-\bar{f}_2^L)$ as a function of $m_h$
for $\tan\beta=50$. 
It has been assumed (according to constraints from $b \to s \gamma$)
that the charged Higgs boson is heavy, $m_{H^\pm}>m_t-m_b$, therefore there
is no absorptive part necessary to develop a non-zero
contribution from diagram f) in Fig.\ref{fig:diag:tbw}.
Additionally, we assumed $C_i = R_{i2}$ (what is a very good approximation
for $\tan\beta=50$) and selected such values for $R_{i2}$
and $R_{i3}$ that are
consistent both with $R_{ij}$ orthogonality and $C_i$ LEP bound,
and provide a maximal value of $\Re(f_2^R-\bar{f}_2^L)$.
}{fig:eett:dec:ref2mhdep}

It is worth to mention that even though in the SM there exists one-loop 
contribution to
$\Re(f_2^R-\bar{f}_2^L)$, 
it turns out to be strongly GIM suppressed~\cite{Grzadkowski:1993gh}.
Therefore, although the 2HDM prediction for $\Re(f_2^R-\bar{f}_2^L)$ 
is smaller than for CP violating from factors in the production mechanism, it
is still by a factor $\sim 10^{5}$ larger then the SM result.

%%%%%%%%%%%%%%%%%%%%%%%%%%%%%%%%%%%%%%%%%%%%%%

\section{Experimental Constraints}
\label{expcon}

Hereafter we will focus on  Higgs boson masses in the region, $m_h=10\div100\gev$.
As it has been shown in the literature~\cite{Gunion:1997aq,*Abreu:2000kg,*Abbiendi:2000ug} 
the existing LEP data are perfectly consistent
with one light Higgs boson within the 2HDM. It turns out that even precision electroweak
tests allow for light Higgs bosons~\cite{Chankowski:1999ta}.

In order to amplify the form factors calculated in this paper we have adopted for
an illustration $\tanb =0.5$. However, there exist experimental constraints on $\tanb$ from
${\rm K^0 - \bar{K}^0}$ and ${\rm B_d-\bar{B}_d}$ mixing~\cite{Gunion:1990vk},
$b\to s \gamma$ decay~\cite{Greub:1999sv} and $Z\to b\bar{b}$ decay~\cite{Haber:1999zh}. 
Since small $\tanb$ enhances $H^\pm t b$ coupling, 
in order to maintain $\tanb =0.5$ we have to decouple charged Higgs effects and therefore
we assume that $m_{H^\pm} \gsim 500\div600\gev$.

The constraints on the mixing angles 
$\alpha_i$  that should be imposed in our numerical analysis are as follows:
\begin{itemize}
\item The $ZZh$ couplings, $C_i^2$,  are restricted by non-observation of
Higgs-strahlung events at LEP1 and LEP2, see Ref.~\cite{Abbiendi:1998rd,*ALEPH2000-028}
\item The contribution to the total $Z$-width from $Z \ra Z^* h_i \ra f
\bar{f} h_i$  is required to be below $7.1 \mev$, see
Ref.~\cite{Ackerstaff:1998ms}.
\end{itemize}
It turns out that the restriction on the $ZZh$ coupling from its contribution to 
the total $Z$-width is always weaker then the one from $Zh$ production if
$m_h \gsim 10\gev$.

The LEP constraints on the $ZZh$ coupling restrict the following entries of the mixing matrix $R_{ij}$:
\begin{equation}
|\sin\beta R_{i2} + \cos_{\beta} R_{i1}|\leq C_i^{exp},
\label{eq:2hdm:LEPZZh}
\end{equation}
where $C_i^{exp}$ stands for the upper limit for the relative strength of $ZZh$ coupling  
determined experimentally in Ref.~\cite{Abbiendi:1998rd,*ALEPH2000-028} up to the Higgs mass $m_h=105\gev$.
As we have concluded in the previous section, CP-violating phenomena we are considering
are enhanced by small $\tan \beta$, in that case one can see from 
eq.(\ref{eq:2hdm:LEPZZh}) that the LEP constraints mostly
restrict $R_{i1}$. Through the orthogonality the restriction on $R_{i1}$  is being transfered to constrain  
$|R_{i2}R_{i3}|=|R_{i2}\sqrt{1-R_{i1}^2-R_{i2}^2}|$ which 
multiplies leading contributions to all CP-violating asymmetries considered here.
The final result for upper limit on $|R_{i2}R_{i3}|$ 
as a function of $\tan \beta$ is shown in Fig.\ref{fig:2hdm:ogrLEPtanb}. In fact the bound on 
$|R_{i2}R_{i3}|$ depends on the Higgs mass, however, in order to be conservative, we have
assumed  $C_i^{exp}=0.12$ that is the most restrictive experimental limit 
(obtained for $\mh\simeq 18\gev$\footnote{ 
For $\mh\simeq 18\gev$ the limits presented 
in Fig.16 of Ref.~\cite{Abbiendi:1998rd,*ALEPH2000-028} 
for  the case when no $b$-tagging  and with $b$-tagging almost coincide.
Therefore our plot in Fig.\ref{fig:2hdm:ogrLEPtanb} is not influenced by
potential problems concerning the dependence of the 
Higgs-$\bbbar$ and Higgs-$\tau^+\tau^-$ branching ratios on the mixing
angles.}).

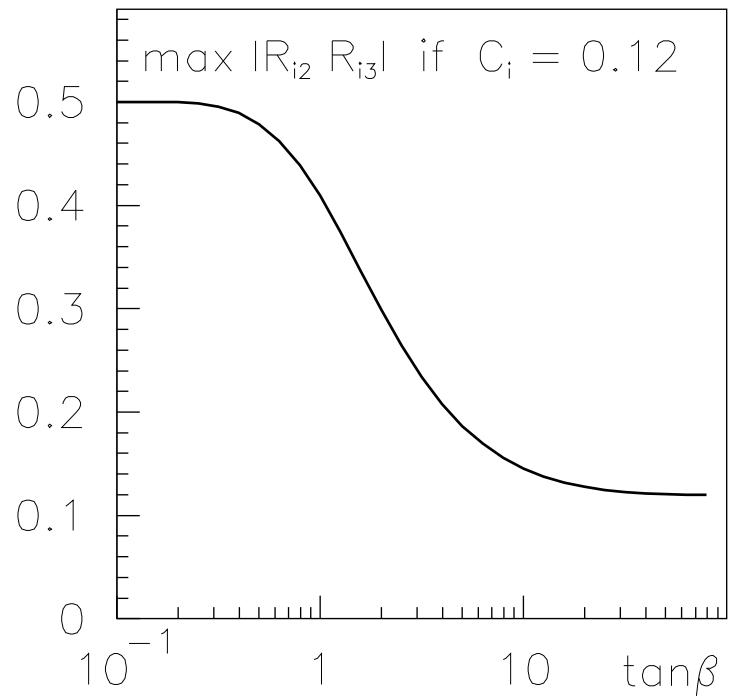
\jpfigure{width=0.65\textwidth}{eett-ogrleptanb.eps}{Maximal value of 
 $|R_{i2}R_{i3}|$ allowed by the LEP constraints on $ZZh_i$ coupling
as a function of $\tb$.}{fig:2hdm:ogrLEPtanb}
 
As it is seen from Fig.\ref{fig:2hdm:ogrLEPtanb} the constraints for 
$|R_{i2}R_{i3}|$
are weak for small $\tanb$. Therefore for $\tan\beta\simeq 0.5 $
it should be legitimate to assume $|R_{i2}R_{i3}|\simeq 1/2$ which is the maximal value consistent 
with orthogonality.

Using the maximal value of $R_{i2}R_{i3}$ allowed by the orthogonality and the LEP constraints
for small $\tanb=0.5$, we may discuss a possibility for an 
experimental determination of the calculated form 
factors at future $\epem$ colliders.
A detailed discussion of expected statistical uncertainties for a measurement of the form factors
has been performed in  Ref.~\cite{Grzadkowski:2000nx}. It has been shown that adjusting
an optimal $\epem$ beam polarizations,
using the energy and angular double distribution of final leptons 
and fitting {\it all 9 form factors} leads
to the following statistical errors for the determination of CP-violating
form factors: $\Delta[\Re(D_\gamma)]=0.08$ and
$\Delta[\Re(D_Z)]=14.4$ for $\epsilon_{\ttbar}\simeq 15\%$.
It is seen that only $\Re(D_\gamma)$, could be
measured with a high precision. We have observed in 
Figs.\ref{fig:eett:prodre:hdep},\ref{fig:eett:prodre:sdep} 
that $\Re(D_\gamma)$ may reach at most a value of $0.10$, therefore one shall conclude that
several years of running with yearly integrated luminosity $L=500\lumun$ should allow for an observation
of $\Re(D_\gamma)$ generated within 2HDM, provided the lightest Higgs boson mass is not too large.
On the other hand, the expected~\cite{Grzadkowski:2000nx} precision for the determination 
of the decay form factors is much more promising:  
$\Delta[\Re(f_2^R-\bar{f}_2^L)]=0.014$. However, as we have seen
in Fig.\ref{fig:eett:dec:ref2mhdep}, the maximal expected size of $\Re(f_2^R-\bar{f}_2^L)$
is  $5 \times 10^{-5}$ (for $m_h > 10\gev$),
therefore either an unrealistic growth of the luminosity, or other observables (besides the energy
and angular double distribution of final leptons)  are required in order to 
observe CP-violating from factors in the top-quark decay process. The results
of Ref.~\cite{Grzadkowski:2000nx} assumed simultaneous\footnote{Obviously, that leads to reduced
precision for the determination of the form factors.} determination of  {\it all 9 form factors}, therefore
another chance to reduce
of $\Delta[\Re(f_2^R-\bar{f}_2^L)]$ is to have some extra independent constraints on the top-quark
coupling coming from other colliders, like the Fermilab Tevatron or LHC.

%%%%%%%%%%%%%%%%%%%%%%%%%%%%%%%%%%%%%%%%%%%%%%%%%%%

\section{CP-Violating Asymmetries}
\label{asymm}
Looking for CP violation one can directly measure~\cite{Grzadkowski:2000nx} 
all the form factors including those which are odd under CP. However another possible 
attitude is to construct  certain asymmetries sensitive to CP violation.
In this section we will discuss several asymmetries that could probe CP violation in the process
$\epem \to \ttbar \to l^\pm \cdots$. We will systematically drop all contributions quadratic in 
non-standard form factors and
calculate various asymmetries keeping only interference between the SM and $D_\gamma$, $D_Z$ 
or $\Re(f_2^R-\bar{f}_2^L)$.

\subsection{Lepton-Energy Asymmetry}
%NPB 484 (1997) 17
Let us introduce the rescaled lepton energy, $x$, by
\beq
x\equiv
\frac{2 E_l}{\mt}\left(\frac{1-\beta_t}{1+\beta_t}\right)^{1/2},
\label{def-x}
\end{equation}
where $E_l$ is the energy of $l$ in $\epem$ c.m. frame and $\beta_t\equiv \sqrt{1-4m_t^2/s}$.
Using lepton energy distribution $d\sigma^\pm/dx$  calculated~\cite{Grzadkowski:1997kn} 
for the general form factors given in eqs.(\ref{vtt},\ref{tbw})  one can define the following 
energy asymmetry:
\begin{equation}
{\cal A}_{CP}^l(x)\equiv\frac{d\sigma^-/dx-d\sigma^+/dx}{d\sigma^-/dx+d\sigma^+/dx}
\label{asy_9604}
\end{equation}
Direct calculation leads to the following result in terms of the CP-violating form factors:
\beq
{\cal A}_{CP}^l(x)=\frac{2\xi g(x) +\Re{(f_2^R-\bar{f}_2^L)}
[\,\delta\! f(x)+\eta\: \delta g(x)\,]}{2\,[\,f(x)+\eta\: g(x)\,]}.
\end{equation}
where
\begin{eqnarray*}
&&\xi \equiv\frac{1}{(3-\beta^2)D_V+2\beta^2 D_A} \\
&&\ \ \ \ \ \times
\frac{-1}{\sin\theta_W}{\Re}\biggl[\:\frac23\: D_\gamma
+\frac{s^2}{(s-m_Z^2)^2}
\frac{(v_e^2+a_e^2)v_t}{64\sin^3\theta_W\cos^3\theta_W}D_Z
\\
&&\ \ \ \ \
-\frac{s}{s-m_Z^2}
\biggl(\,\frac{v_e v_t}{16\sin^2\theta_W\cos^2\theta_W}D_\gamma
+\frac{v_e}{6\sin\theta_W\cos\theta_W}D_Z \,\biggr)\:\biggr],
\end{eqnarray*}
for
\begin{eqnarray*}
&&D_V=(v_e v_t d-\frac23)^2 +(a_e v_t d)^2, \\
&&D_A=(v_e a_t d)^2 +(a_e a_t d)^2,         \\
\end{eqnarray*}
with the SM neutral-current parameters of $e$ and
$t$: $v_e=-1+4\sin^2\theta_W$, $a_e=-1$, $v_t=1-(8/3)\sin^2\theta_W$,
and $a_t=1$, and a $Z$-propagator factor
$$
d\equiv\frac{s}{s-m_Z^2}
\frac{1}{16\sin^2\theta_W\cos^2\theta_W}.
$$
The coefficient $\eta$ is defined as
$$
\eta\equiv
\frac{{4\:{\Re}(D_{V\!A})}}{{(3-\beta^2)D_V +2\beta^2 D_A}}.
$$
for 
$$
D_{V\!A}=v_e a_t d(v_e v_t d-\frac23) +(a_e d)^2 a_t v_t.
$$
The definitions of the functions $f$, $g$, $\delta f$ and $\delta g$ 
could be obtained from Ref.~\cite{Grzadkowski:1997kn}.
\nc{\ggttbar}{g_{\gamma \ttbar}}
\nc{\gZttbar}{g_{Z \ttbar}}
\nc{\gWtb}{g_{Wtb}}

\jpfigure{width=.65\textwidth}{eett-asy9604ff-xdep.eps}{
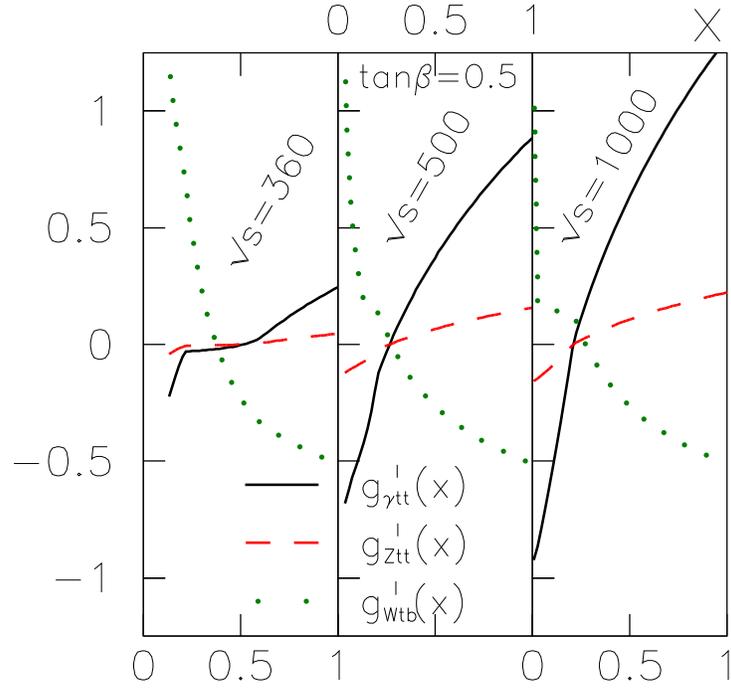
The coefficient functions $g$ defined by eq.(\ref{eq:asym1:decomposition}) for
 the energy-asymmetry 
as function of $x$ for $\sqrt{s}=$ 360 (left), 500 (middle)
and 1000 GeV (right) for $\tanb=0.5$. 
The solid curve represents the coefficient $\ggttbar^l(x)$,
dashed $\gZttbar^l(x)$ and dotted 
$\gWtb^l(x)$.}{fig:eett:asy9604f:xdep}

In order to estimate a relative strength of various sources of CP violation 
it is worth to decompose the asymmetry as follows:
\begin{equation}
{\cal A}_{CP}^l(x)=\ggttbar^l(x)\ \ReDgamma+ 
                   \gZttbar^l(x)\ \ReDZ+
                   \gWtb^l(x)   \ \Refrmfl.
\label{eq:asym1:decomposition}
\end{equation}
As one can see from Fig.\ref{fig:eett:asy9604f:xdep} the CP-violating effects that originate 
from the decay
are substantially enhanced in the soft energy region. 
It is worth to notice that the minimal lepton energy for $\sqrt{s}=500\gev$ is $E_l^{min}\simeq7.5\gev$ that
is large enough to detect the lepton. 
Therefore the region of soft leptons should be carefully studied experimentally.
The enhancement is a consequence of particular 
behavior of $f(x)+\eta g(x)$, $g(x)$ and $\delta f(x)+\eta \delta g(x)$ that causes the relative
amplification of the decay effects, see Fig.1 in Ref.~\cite{Brzezinski:1997av}.
The same figure explains the observed smallness of the $Z$-boson contribution.
For hard leptons both $\Refrmfl$  and $\ReDgamma$ are enhanced and they are
raising with the c.m. energy.

The energy-asymmetry could be decomposed into the leading contribution proportional to
$R_{i2}R_{i3}$ and the remaining piece proportional to $R_{i1}R_{i3}$. The former one 
(that provides the leading contribution) is 
plotted in Fig.\ref{fig:eett:asy9604r:mhdep} for a fixed energy, $x=0.8$, as a function of $m_h$.
One can observe
that the largest asymmetry for the chosen energy corresponds to $\sqrt{s}=360\gev$ and 
$m_h=10\gev$.

\jpfigure{width=.65\textwidth}{eett-asy9604r-mhdep.eps}{The Higgs mass dependence of 
the coefficient  of $R_{i2} R_{i3}$ for the asymmetry given by eq.(\ref{asy_9604})
for $\sqrt{s}$=360 (solid), 500 (dashed), 
1000 GeV (dotted) for a fixed energy, $x_l=0.8$, and
$\tanb=0.5$. }{fig:eett:asy9604r:mhdep}

\subsection{Integrated Lepton-Energy Asymmetry}
%PLB 391 (1997) 172
CP symmetry could be also tested using the following 
leptonic double energy distribution~\cite{Grzadkowski:1997pc}:
\beq
\frac{1}{\sigma}\frac{d^2\sigma}{dx\;d\barx}
=\sum_{i=1}^{3}c_i f_i(x,\barx),
\label{DD}
\end{equation}
where $x$ and $\barx$ are for $l^+$ and $l^-$ respectively, for
$$
c_1=1,\ \ \ 
c_2=\xi,\ \ \ 
c_3=\frac{1}{2}\Re(f_2^R-\bar{f}_2^L)
$$
and
\begin{eqnarray}
&&f_1(x,\bar{x})= f(x)f(\barx)+\eta'g(x)g(\barx)
+\eta[f(x)g(\barx)+g(x)f(\barx)], \non \\ 
&&f_2(x,\bar{x})= f(x)g(\bar{x})-g(x)f(\bar{x}), \non \\ 
&&f_3(x,\bar{x})=\delta\! f(x)f(\bar{x})-f(x)\delta\! f(\bar{x})
+\eta'[\delta g(x)g(\bar{x})-g(x)\delta g(\bar{x})] \non \\
&&\ \ \ \ \ \ \ \ \ \ \
+\eta[\delta\! f(x)g(\bar{x})-f(x)\delta g(\bar{x})
+\delta g(x)f(\bar{x})-g(x)\delta\! f(\bar{x})], \non
\end{eqnarray}
where 
$$
\eta'\equiv\frac{1}{\beta^2}\frac{(1+\beta^2)D_V+2\beta^2 D_A}
{(3-\beta^2)D_V+2\beta^2 D_A}.
$$

The following asymmetry could be a measure of CP violation:
\begin{equation}
A_{CP}^{ll}\equiv
\frac
{\dps\int\int_{x<\bar{x}}dxd\bar{x}\frac{d^2\sigma}{\dps dxd\bar{x}}
 -\int\int_{x>\bar{x}}dxd\bar{x}\frac{d^2\sigma}{\dps dxd\bar{x}}}
{\dps\int\int_{x<\bar{x}}dxd\bar{x}\frac{d^2\sigma}{\dps dxd\bar{x}}
 +\int\int_{x>\bar{x}}dxd\bar{x}\frac{d^2\sigma}{\dps dxd\bar{x}}}.
\label{asy_9608}
\end{equation}
As before, it is useful to separate contributions from various form factors:
\begin{equation}
A_{CP}^{ll}=\ggttbar^{ll}\ \ReDgamma+
            \gZttbar^{ll}\ \ReDZ+
            \gWtb^{ll}   \ \Refrmfl.
\label{h_def}
\end{equation}
In Table~\ref{s_dep_ff} we show the coefficients $g$ for various c.m. energies.
Firstly, is clear that for any given $\sqrt{s}$ the coefficient $\gZttbar^{ll}$ is the smallest one.
Secondly, it is seen that just above the threshold for $\ttbar$ production there is an enhancement of
relative contributions from the decay, however that still not sufficient to overcome the suppression
of $\Re(f^R_2-\bar{f}^L_2) $ that we have observed in Fig.\ref{fig:eett:dec:ref2mhdep}.
Therefore we can conclude  that the leading contribution is provided by CP violation in 
the $\gamma\ttbar$ vertex. 

\vspace{1cm}
\begin{table}[h]
\begin{center}
\begin{tabular}{|r|r @{.} l|r @{.} l|r @{.} l|}
\hline
$\sqrt{s}$[GeV]& 
\multicolumn{2}{c|}{$\ggttbar^{ll}$} & 
\multicolumn{2}{c|}{$\gZttbar^{ll}$}&
\multicolumn{2}{c|}{$\gWtb^{ll}$}\\
\hline
360&0&0509 & 0&00954 & 0&410 \\
\hline
500&0&386 & 0&0684 & 0&291 \\
\hline
1000&0&602 & 0&102 & 0&235 \\
\hline
\end{tabular}
\end{center}
\caption{The energy dependence of the coefficients $g$ defined in eq.(\ref{h_def}).}
\label{s_dep_ff}
\vspace*{0.5cm}
\end{table}

Fig.\ref{fig:eett:asy9608r:mhdep} illustrates the Higgs-mass dependence of the leading
(proportional to $R_{i2} R_{i3}$) contribution to the integrated lepton-energy asymmetry.
It turns out that $\sqrt{s}=500\gev$ provides the largest asymmetry.

\jpfigure{width=.65\textwidth}{eett-asy9608r-mhdep.eps}{
Higgs mass dependence of the coefficient of
$R_{i2} R_{i3}$  for the asymmetry given by eq.(\ref{asy_9608})
for $\sqrt{s}$=360 (solid), 500 (dashed), 1000 GeV (dotted) for 
$\tanb=0.5$. }{fig:eett:asy9608r:mhdep}

Using results of Ref.~\cite{Grzadkowski:1997pc} one can find out an expected
statistical error for the determination of $A_{CP}^{ll}$ at any given $\epem$ collider.
Assuming $\sqrt{s}=500\gev$, $L=500\lumun$ and lepton tagging efficiency, 
$\epsilon_l=60\%$ we get $\Delta A_{CP}^{ll}=0.014$. 
As it is seen from Fig.\ref{fig:eett:asy9608r:mhdep}
an observation of the asymmetry would require several years of running
at the assumed luminosity.

\subsection{Angular Asymmetry}
% PLB 476 (2000) 87

Another CP-violating asymmetry could be constructed using
the angular distributions of the bottom quarks or leptons originating from 
the top-quark decay:
\beq
\frac{d\sigma}{d\cos\thf}=
\frac{3\pi\beta\alpha_{\mbox{\tiny EM}}^2}{2s}B_{\sst{f}}
\left({\mit\Omega}_0^{\sst{f}}+{\mit\Omega}_1^{\sst{f}}
\cos\thf+{\mit\Omega}_2^{\sst{f}}\cos^2\thf\right),   
\label{dis2}
\end{equation}
where $f=b,l$, $B_f$ is an appropriate 
top-quark branching ratio, $\theta_f$ is the angle between the $e^-$ beam direction and the
direction of $f$ momentum in the $\epem$ c.m. frame and $\Omega_i^f$ are coefficients
calculable in terms of the form factors, see Ref.~\cite{Grzadkowski:1999iq}.
The following asymmetry provides a signal of CP violation:
\beq
{\cal A}_{CP}^f(\thf)= \frac{
{\displaystyle \frac{d\sigma^+(\thf)}{d\cos\thf}-
\frac{d\sigma^-(\pi-\thf)}{d\cos\thf}}}
{{\displaystyle \frac{d\sigma^+(\thf)}{d\cos\thf}+
\frac{d\sigma^-(\pi-\thf)}{d\cos\thf}}},
\label{asy_9911}
\end{equation}
where $d\sigma^{+/-}$ is referring to $f$ and $\bar{f}$
distributions, respectively. Since $\theta_f \to \pi-\theta_{\bar{f}}$
under $CP$, the asymmetry defined above is a true measure of $CP$
violation.

Adopting general formulas for the asymmetry from Ref.~\cite{Grzadkowski:1999iq} 
and inserting form factors calculated here we plot the asymmetry in 
Figs.\ref{fig:eett:asy9911bff-thdep},~\ref{fig:eett:asy9911lff-thdep}
as a function of $\cos\theta$ for bottom quarks and leptons, respectively.
As before, the asymmetry can be decomposed into $\gamma\ttbar$, $Z\ttbar$ and $Wtb$
vertex contributions:
\begin{equation}
{\cal A}_{CP}^f(\thf)=\ggttbar^f(\thf)\ \ReDgamma+ 
                   \gZttbar^f(\thf)\ \ReDZ+
                   \gWtb^f(\thf)   \ \Refrmfl.
\label{gAf}
\end{equation}
It is seen that forward-backward directions are favored, however an experimental cut
$|\cos\theta_f| < 0.9$ should be imposed in the realistic  experimental environment.

\jpfigure{width=.65\textwidth}{eett-asy9911bff-thdep.eps}{
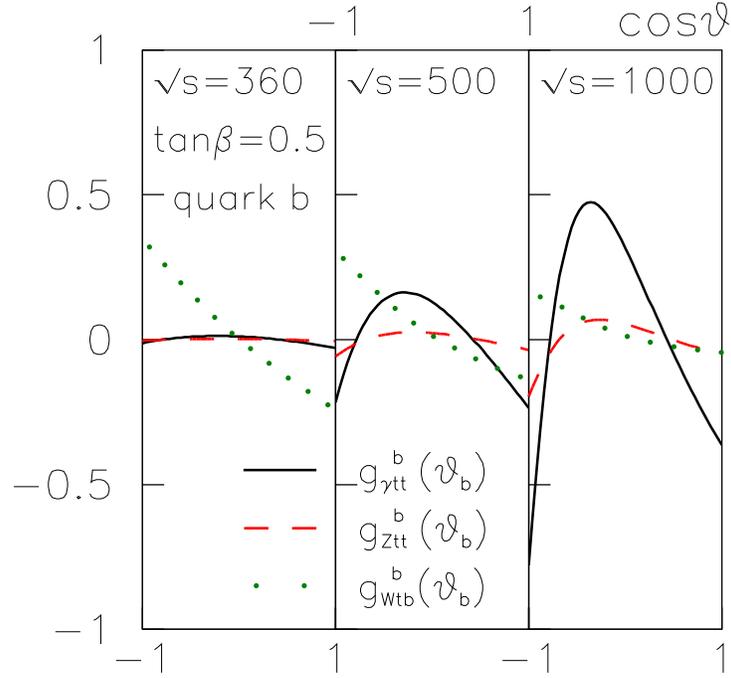
The coefficients functions $g$ defined by eq.(\ref{gAf}) for the angular asymmetry, 
 for final bottom quarks ($f=b$) 
as a function of 
$\cos\theta$ for $\sqrt{s}=$ 360 (left), 500 (middle) and 1000 GeV (right) 
for $\tanb=0.5$. The solid curve represents the coefficient  $\ggttbar^f(\thf)$,
dashed $\gZttbar^f(\thf)$ and dotted  $\gWtb^f(\thf)$}{fig:eett:asy9911bff-thdep}

\jpfigure{width=.65\textwidth}{eett-asy9911lff-thdep.eps}{
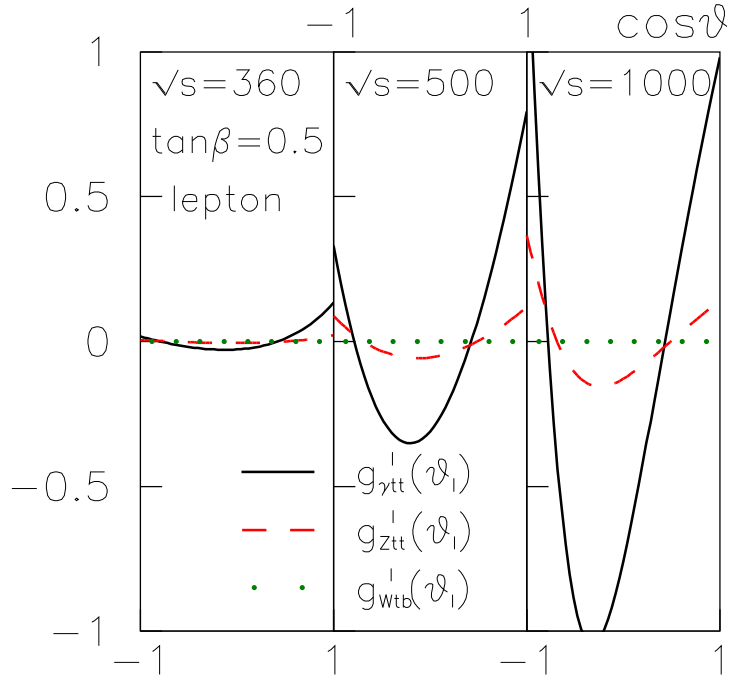
The same as Fig.\ref{fig:eett:asy9911bff-thdep} calculated for
final leptons ($f=l$).}{fig:eett:asy9911lff-thdep}

In order to illustrate the Higgs mass dependence we plot in Fig.\ref{fig:eett:asy9911br-mhdep}
the angular asymmetry both for $f=b$ and $f=l$ for chosen polar angle $\cos \theta=-0.25$.
As expected the maximal effect could be reached for minimal Higgs mass $m_h=10\gev$, 
$\sqrt{s}=500\gev$ is the most suitable energy.

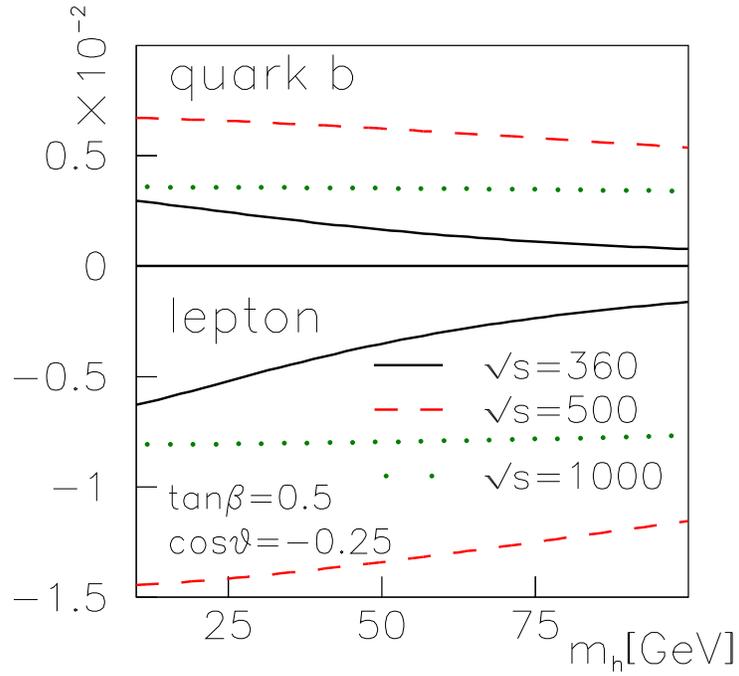
\jpfigure{width=.65\textwidth}{eett-asy9911r-mhdep.eps}{The Higgs mass dependence of 
the coefficient  of $R_{i2} R_{i3}$ for the angular asymmetry defined by eq.(\ref{asy_9911})
for $\sqrt{s}$=360 (solid), 500 (dashed), 1000 (dotted) GeV for a fixed polar angle
$\cos \theta=-0.25$ and $\tanb=0.5$.}{fig:eett:asy9911br-mhdep}

\subsection{Integrated Angular Asymmetry}
% NPB585 (2000) 3
The angular distribution given in eq.(\ref{dis2}) could be adopted to define an integrated 
version~\cite{Grzadkowski:2000nx}
of the angular asymmetry ${\cal A}_{CP}^f(\thf)$:
\begin{equation}
{\cal A}_{\sst{CP}}^{\ssf}(P_{e^-},P_{e^+})= \frac{
{\displaystyle \int_{-c_m}^{0}\!d\cos\thf
 \frac{d\sigma^{+(*)}(\thf)}{d\cos\thf}
 -\int_{0}^{+c_m}\!d\cos\thf \frac{d\sigma^{-(*)}(\thf)}{d\cos\thf}}}
{{\displaystyle \int_{-c_m}^{0}\!d\cos\thf
  \frac{d\sigma^{+(*)}(\thf)}{d\cos\thf}
 +\int_{0}^{+c_m}\!d\cos\thf \frac{d\sigma^{-(*)}(\thf)}{d\cos\thf}}}, 
\label{asy_0004}
\end{equation}
where $P_{e^-}$ and $P_{e^+}$ are the polarizations of $e$ and
$\bar{e}$ beams, $d\sigma^{+/-(*)}$ is referring to $f$ and $\bar{f}$
distributions respectively, and $c_m$ expresses the experimental
polar-angle cut. In order to discuss possible advantages of polarized initial beams
we are considering here dependence of the asymmetry on
the polarization. Hereafter we will discuss the same polarization for
$e$ and $\bar{e}$: $P\equiv P_{e^-}=P_{e^+}$. 
 
Again we decompose the asymmetry as follows:
\begin{equation}
{\cal A}_{CP}^f(P)=\ggttbar^f(P)\ \ReDgamma+ 
                   \gZttbar^f(P)\ \ReDZ+
                   \gWtb^f(P)   \ \Refrmfl.
\label{g_def}
\end{equation}
In Table~\ref{s_dep_g} we show the coefficient functions $g$ calculated for various energy
and polarization choices assuming the polar angle cut $|\cos\thf|<
0.9$, i.e. $c_m=0.9$ in eq.(\ref{asy_0004}), both for leptons and bottom
quarks\footnote{Note that in Table~\ref{s_dep_g} there is no column corresponding to the coefficient of 
$\Re(f^R_2-\bar{f}^L_2)$. That happens since the angular distribution for leptons is not 
influenced by corrections to the top-quark decay vertex, see 
Refs.~\cite{Grzadkowski:1999iq,Rindani:2000jg} 
and~\cite{Grzadkowski:2000nx}.} . It could be seen that a positive polarization leads to 
higher coefficients $g_{\gamma\ttbar}^f$ and
$g_{Z\ttbar}^f$.
Since $\Re(D_\gamma) > \Re(D_Z) \gg \Re(f^R_2-\bar{f}^L_2) $ that implies that 
maximal asymmetry could be reached for $P=+0.8$ and 
the dominant contribution is originating from $\Re(D_\gamma)$.
Since the number of events does not drop drastically when going from unpolarized beams to $P=+0.8$,
it turns out that the positive polarization is the most suitable for testing the integrated angular asymmetry.
It is clear from the table that the asymmetry for final leptons should be larger by a factor $3\div4$ then
the one for bottom quarks and their signs should be reversed.

\begin{table}
\begin{center}
\begin{tabular}{|r|r|r @{.} l|r @{.} l|r @{.} l|r @{.} l|r @{.} l|}
\hline
$\sqrt{s}[\gev]$&$P$&\multicolumn{6}{c|}{quark b}&\multicolumn{4}{c|}{lepton}\\
\cline{3-12}
&&
\multicolumn{2}{c|}{$g_{\gamma\ttbar}^b(P)$} & 
\multicolumn{2}{c|}{$g_{Z\ttbar}^b(P)$}&
\multicolumn{2}{c|}{$g_{Wtb}^b(P)$}&
\multicolumn{2}{c|}{$g_{\gamma\ttbar}^l(P)$} & 
\multicolumn{2}{c|}{$g_{Z\ttbar}^l(P)$}\\ 
\hline
360 & 0.0& 0&00844& 0&00106& 0&142& -0&0162&-0&00203\\ 
    & 0.8& 0&00983&-0&00555&-0&259& -0&0493& 0&0278\\
    &-0.8& 0&00758& 0&00510& 0&388& -0&0106&-0&00713\\
\hline 
500 & 0.0& 0&113&   0&0136&  0&121& -0&224& -0&0270\\
    & 0.8& 0&131&  -0&0718& -0&247& -0&627&  0&343\\
    &-0.8& 0&101&   0&0661&  0&347& -0&149& -0&0968\\
\hline
1000& 0.0& 0&332&   0&0389&  0&0678&-0&722& -0&0845\\
    & 0.8& 0&422&  -0&225&  -0&167& -1&55&   0&824\\
    &-0.8& 0&284&   0&181&   0&194& -0&507& -0&322\\
\hline
\end{tabular} 
\end{center}
\caption{The energy and polarization dependence of the coefficients $g_{\gamma\ttbar}^f(P)$,
$g_{Z\ttbar}^f(P)$ and  $g_{Wtb}^f(P)$ defined in eq.(\ref{g_def}) for leptons ($f=l$) 
and bottom quarks ($f=b$).}
\label{s_dep_g}
\end{table}

Using the general formula for the asymmetry 
from Ref.~\cite{Grzadkowski:2000nx} and adopting results for the CP-violating form factors
we plot ${\cal A}_{\sst{CP}}^{\ssf}(P_{e^-},P_{e^+})$ in Fig.\ref{fig:eett:asy0004br-mhdep}
as a function of the Higgs mass both for bottom quarks and leptons. It is clear that the largest 
asymmetry could be expected for $P_{e^-}=P_{e^+}=+0.8$ for final leptons at $\sqrt{s}=500\gev$.
With the maximal mixing, $R_{i2}R_{i3}=1/2$ the $1\%$ asymmetry could be expected for
the Higgs boson with mass $m_h=10\div20\gev$. Since the statistical error 
expected~\cite{Grzadkowski:2000nx} for
the asymmetry is of the order of $5\times 10^{-3}$, we can conclude that the asymmetry
${\cal A}_{\sst{CP}}^{\ssf}(P_{e^-},P_{e^+})$ is the most promising one, leading to
$2\sigma$ effect for light Higgs mass and $\tanb=0.5$. 
As it is seen form Fig.\ref{fig:eett:asy0004br-mhdep}
it is relevant to have polarized $\epem$ beams.

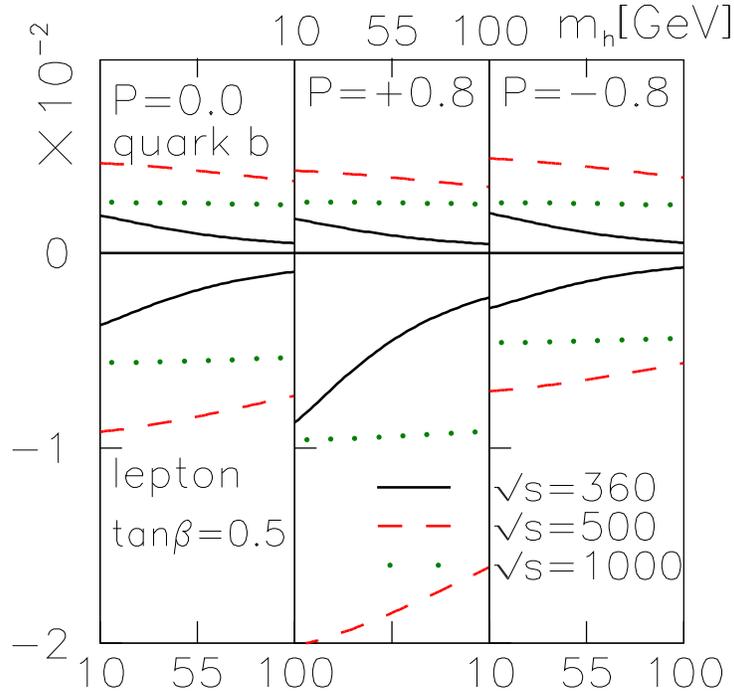
\jpfigure{width=.65\textwidth}{eett-asy0004r-mhdep.eps}{The Higgs mass dependence of 
the coefficient  of $R_{i2} R_{i3}$ for the angular asymmetry defined by eq.(\ref{asy_0004})
for bottom quarks (upper) and leptons (lower) at $\sqrt{s}$=360 (solid), 500 (dashed), 
1000 GeV (dotted) with unpolarized beams (left), $P=+0.8$ (middle) and $P=-0.8$ (right)
for $\tanb=0.5$.}{fig:eett:asy0004br-mhdep}

%%%%%%%%%%%%%%%%%%%%%%%%%%%%%%%%%%%%%%%%%%%%%%%%%%%

\section{Summary and Conclusions}
\label{summary}

We have considered a general two-Higgs-doublet model with CP violation in the scalar sector.
Mixing of the three neutral Higgs fields of the model leads to CP-violating Yukawa couplings
of the physical Higgs bosons. CP-asymmetric form factors generated at the one-loop level of 
perturbation theory has been calculated within the model.
Although in general the existing experimental data from LEP1 and LEP2 constraint
the mixing angles of  the three neutral Higgs fields, their combination relevant
for CP violation is not bounded for small $\tanb$ which is the region of our interest.
We have shown that the decay form factors are typically smaller then the production ones
by 2-3 orders of magnitude. The dominant contribution to CP violation in the production is coming from
$\gamma\ttbar$ coupling. Several energy and angular
CP-violating asymmetries for the process
$\epem \to \ttbar \to l^\pm \cdots$ and $\epem \to \ttbar \to \bb \cdots$ has been considered
using the form factors calculated within the two-Higgs-doublet model.
It turned out that the best test of CP invariance would be provided by the integrated 
angular asymmetry ${\cal A}_{\sst{CP}}^{\ssf}(P_{e^-},P_{e^+})$ for positive 
polarizations of $\epem$ beams. For one year of running at TESLA collider with the integrated luminosity
$L=500\lumun$ one could  expect $2\sigma$ effect  for the asymmetry for light Higgs boson
and $\tanb=0.5$.

\vspace{1.5cm}
\centerline{\bf Acknowledgments}
\vspace{.5cm} This work was supported in part by the Committee for
Scientific Research (Poland) under grants No.~2~P03B~014~14 and No.~2~P03B~080~19.

\vspace{1cm}

\def\thesection{\Alph{section}}
\setcounter{section}{0}%
\section{Appendix}
\subsection{1-Loop Integrals}
\label{app-loops}
    
\jpfigure{width=0.5\textwidth}{phd-diag-general-triangle.eps}{
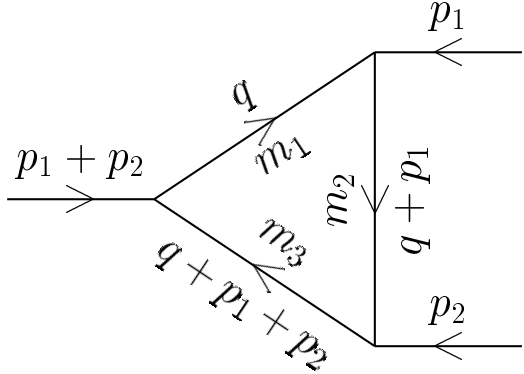
Convention for momenta and mass labelling used.}{fig:diag:general:triangle}

Here we will define 3-point one-loop integrals used in calculations of form factors.
The convention for momenta and mass labelling is presented in Fig.\ref{fig:diag:general:triangle}.
The Passarino-Veltman functions~\cite{PassarinoVeltman} 
$\loopC{}{}$ are defined as follows:
\begin{eqnarray}
&\mu^{4-n}& \int \frac{d^n k}{(2\pi)^n} \frac{1; k_{\mu}; k_{\mu}k_{\nu}}
{[k^2 - m_1^2][(k + p_1)^2 - m_2^2][(k + p_1 + p_2)^2 - m_3^2]} 
  \nonumber \\
& & = \frac{i}{16\pi^2} 
\loopC{0; \mu; \mu\nu}{}(p_1^2, p_2^2, p^2; m_1^2, m_2^2, m_3^2).
        \label{eqn:3pointloopintdef}
\end{eqnarray}
The vector and tensor integrals $\loopC{\mu}{}$ and 
$\loopC{\mu\nu}{}$ can be
expanded into scalar coefficients and Lorentz covariants:
\begin{eqnarray}
\loopC{}{\mu} &=& p_1^{\mu} \loopC{11}{} + p_2^{\mu} \loopC{12}{} \\
\loopC{}{\mu\nu} &=&  p_1^{\mu} p_1^{\nu} \loopC{21}{} 
   + p_2^{\mu} p_2^{\nu} \loopC{22}{}
    + (p_1^{\mu} p_2^{\nu} + 
    p_2^{\mu} p_1^{\nu}) \loopC{23}{}+ g^{\mu\nu} \loopC{24}{} .
\end{eqnarray}

Imaginary part of $\loopC{12}{}$ for certain sets of arguments 
is needed for the calculation
of $\Re D_\gamma$ and $\Re D_Z$:
\begin{eqnarray}
&&\Im \loopC{12}{}(p_t,p_{\bar{t}};m_t,m_h,m_t)
=\frac{\pi}{s\beta_t} \left[
  1-\frac{h^2}{\beta_t^2} \log \left(
    1+\frac{\beta_t^2}{h^2}
  \right)
\right] \Theta(\sqrt{s}-2 m_t)\non \\
&&\Im \loopC{12}{}(p_t,p_{\bar{t}};m_h,m_t,m_Z)
=\frac{\pi}{s\beta_t^2} \times \\ 
&&\left\{\left[2 t^2 (1-h^2-z^2) -h^2 \beta_t^2 \right]\frac{1}{\beta_t}\log\left(
     \frac{1-z^2-h^2-\beta_t\beta_Z}{1-z^2-h^2+\beta_t\beta_Z}
   \right) + \beta_Z
\right\}\Theta(\sqrt{s}-m_h-m_Z), \non
\end{eqnarray}

for $h\equiv m_h/\sqrt{s}$, $\beta_t=\sqrt{1-4m_t^2/s}$ 
and $\beta_Z=\lambda(1,z^2,h^2)$, where $\lambda(a,b,c)$ is the standard
kinematic function.

In order to calculate $\re(f^R_2-\bar{f}^L_2) $ one needs an 
imaginary part of the following $\loopC{22}{}$ function.
Neglecting $m_b$ and defining $\hat{m}_W\equiv\frac{m_W}{\omega m_t}$ and 
$\hat{m}_h\equiv\frac{m_h}{\omega m_t}$
for $\omega=1-m_W^2/m_t^2$ one gets:
\nc{\mhhs}{\hat{m}_h^2}
\nc{\mwhs}{\hat{m}_W^2}
\bea
&&\Im \loopC{22}{}(p_W,  p_b, m_W^2, m_h^2, m_b^2)=\frac{\pi\omega}{2 m_t^2}\left\{ \left[
  2 \mhhs ( 6 \mwhs + 1) -  (2 \mwhs  + 1)\right] +
\right.\nnr\\
&&\left.
-  2 \left[ \mhhs ( 6 \mwhs + 1) + 2 \mwhs  \right] \mhhs \log\left(1 + \hat{m}_h^{-2}\right)
\right\}.
\end{eqnarray}

\subsection{Asymptotic Formulae}
\label{app-asymptotic}

\subsubsection{Production of $\ttbar$ }
\newcommand{\mhredp}{\hat{m}_h^2}
For $h\equiv \frac{m_h}{\sqrt{s}}$ in the limit $\frac{h^2}{\beta_t^2} \ll 1$ one gets:
\begin{equation}
\Re D_{\gamma}=-\frac{t^2 }{2  \pi \beta_t } \frac{m_t^2}{v^2} A_\gamma  b_t a_t
\left[1  + \frac{h^2}{\beta_t^2} \log\left(\frac{h^2}{\beta_t^2}\right)\right] \Theta(1-4 t^2)
\end{equation}
\begin{multline}
\Re D_Z= -\frac{1}{2 \pi \beta_t^3 } \frac{m_t^2}{v^2} A_Z b_t  
\left\{
a_t t^2 \beta_t^2 \left[1  + \frac{h^2}{\beta_t^2} \log\left(\frac{h^2}{\beta_t^2}\right)\right]
\Theta(1- 4 t^2)
+\right.\\
-C_i z^2 \left[
     \beta_t (1 - h^2 - z^2 ) 
     + ( 2 t^2 (1 -  h^2 - z^2) -  h^2 \beta_t^2 ) 
      \log\left(\frac{1-\beta_t}{1 + \beta_t}\right)\right] \\
\left.
\Theta{}(1-h-z),
\right\}
\label{small_prod}
\end{multline}
where $z\equiv m_Z/\sqrt{s}$ and $t\equiv \mt/\sqrt{s}$.

For $\frac{h^2}{\beta_t^2 s} \gg 1$ one gets:
\bea
\Re D_{\gamma}&=&-\frac{ t^2}{12  \pi \beta_t}\frac{m_t^2}{v^2} A_\gamma  b_t a_t 
\frac{\beta_t^2}{h^2} (3 - 2\frac{\beta_t^2}{h^2} ) \Theta(1-4 t^2)\non \\
\Re D_Z&=&-\frac{t^2 }{12 \pi \beta_t } \frac{m_t^2}{v^2} A_Z b_t a_t 
\frac{\beta_t^2}{h^2} (3 - 2\frac{\beta_t^2}{h^2} ) \Theta(1- 4 t^2)
\eea

\subsubsection{Decay of the Top Quark}
\newcommand{\mhred}{\hat{m}^2_h}
In the limit $\hat{m}_h\equiv\frac{m_h}{\omega m_t} \ll 1$:
\begin{equation}
\Re(f_2^R|_{CPV})=\frac{g^2 b_b C_i}{64 \pi}  
\left(\frac{m_b}{ m_t}\right)^2 \hat{m}_W
\left\{
- (1 + \frac{1}{2} \hat{m}_W^{-2} ) 
+ \left[6 + \hat{m}_W^{-2} + 2 \log(\mhred)\right] \mhred 
\right\},
\end{equation}
while for  $\hat{m}_h\gg 1$:
\begin{equation}
\Re(f_2^R|_{CPV})= \frac{g^2 b_b C_i}{64 \pi} \left(\frac{m_b}{ m_t}\right)^2 
\hat{m}_W^{-1} \hat{m}_h^{-2} \left[
- (\hat{m}_W^2 + \frac{1}{3} ) 
+ (\frac{5}{6}\hat{m}_W^2 + \frac{1}{4}) \hat{m}^{-2}_h
\right].
\end{equation}

\bibliographystyle{phcpc_mcite}
\bibliography{ref}
\end{document}